\documentclass{jetpl}
\usepackage{cite}
\twocolumn

\lat

\title{Direct Evidence of Two Superconducting Gaps in FeSe$_{0.5}$Te$_{0.5}$: SnS-Andreev Spectroscopy and Lower Critical Field}

\rtitle{Direct Evidence of Two superconducting Gaps in FeSe$_{0.5}$Te$_{0.5}$}

\sodtitle{Direct Evidence of Two superconducting Gaps in FeSe$_{0.5}$Te$_{0.5}$: SnS-Andreev Spectroscopy and Lower Critical Field}
\author{T.E. Kuzmicheva$^{\dag}$\/\thanks{e-mail: kute@sci.lebedev.ru},
S.A. Kuzmichev$^{\ast,\dag}$,
A.V. Sadakov$^{\dag}$,
A.V. Muratov$^{\dag}$,
A.S. Usoltsev$^{\dag}$,
V.P. Martovitsky$^{\dag}$,
A.R. Shipilov$^{\ast}$,
D.A. Chareev$^{\ddag, +, \diamond}$,
E.S. Mitrofanova$^{\ddag}$,
and V.M. Pudalov$^{\dag}$}

\rauthor{T.E. Kuzmicheva, S.A. Kuzmichev, A.V. Muratov, et al.}

\sodauthor{Kuzmicheva, Kuzmichev, Sadakov, Muratov, Usoltsev, Martovitsky, Shipilov, Chareev, Mitrofanova, Pudalov}

\address{$^{\dag}$P.N. Lebedev Physical Institute RAS,
119991 Moscow, Russia\\~\\
$^{\ast}$M.V. Lomonosov Moscow State University, 119991 Moscow, Russia\\~\\
$^{\ddag}$Institute of Experimental Mineralogy RAS, 142432 Chernogolovka, Russia \\~\\
$^+$ Institute of Physics and Technology, Ural Federal University, 620002 Ekaterinburg, Russia \\~\\
$^{\diamond}$ Kazan Federal University, 420008 Kazan, Russia}

\date{\today}

\abstract{We present direct measurements of the superconducting order parameter in nearly optimal FeSe$_{0.5}$Te$_{0.5}$ single crystals with critical temperature $T_C \approx 14$\,K. Using intrinsic multiple Andreev reflection effect (IMARE) spectroscopy and measurements of lower critical field, we directly determined two superconducting gaps, $\Delta_L \approx 3.3 - 3.4$\,meV and $\Delta_S \approx 1$\,meV, and their temperature dependences. We show that a two-band model fits well the experimental data. The estimated electron-boson coupling constants indicate a strong intraband and a moderate interband interaction.}
\PACS{74.20.Mn,74.25.-q,74.25.Ha,74.45.+c,74.70.Xa}

\begin{document}

\maketitle

\section{Introduction}
Among other Fe-based superconductors, iron selenide has the simplest crystal structure of FeSe layers weakly connected along the $c$-direction. Its critical temperature, rather low in stoichiometric state, $T_C \approx 9 - 11$\,K, could be substantially raised up to 15\,K \cite{Chareev1,Chareev2} by Te substitution, or up to 37\,K by applied pressure \cite{Margadonna}, and reaches as high as 100\,K in FeSe monolayer on SrTiO$_3$ substrate \cite{Ge} (for a review, see \cite{Wirth}). Band-structure calculations \cite{Subedi} showed several bands crossing Fermi level, and cylinder-like Fermi surface sheets, electron around M point, and hole around $\Gamma$ point. At temperatures below $T_C$, two superconducting condensates are seemed to develop.

Nonetheless, despite intensive eight-year research, theoretical studies of the underlying pairing mechanism in Fe-based pnictides and selenides are far from consensus. Several competing models were developed, so called $s^{\pm}$ \cite{Maiti}, $s^{++}$ \cite{Kontani}, and superstripe model \cite{Bianconi}, but no one has got an unambiguous experimental confirmation yet. The available experimental data with Te-substituted iron selenide are rather contradictory. Single superconducting gap with BCS-ratio $2.9 - 7$ was observed in \cite{Park,Kato,Wu,Nakayama}, whereas point-contact Andreev reflection (PCAR) \cite{Daghero}, muon-spin-rotation ($\mu$SR) \cite{Khasanov}, tunneling \cite{Hanaguri}, angle-resolved photoemission spectroscopy (ARPES) \cite{Miao}, specific heat \cite{Hu}, and scanning tunneling microscopy (STM) \cite{Yin} measurements resolved two distinct gaps with $2\Delta_L/k_BT_C = 3.6 - 7.2$ and $2\Delta_S/k_BT_C = 0.8 - 4$. Using PCAR technique, Daghero et al. \cite{Daghero} studied superconducting order parameter for FeSe$_{1-x}$Te$_x$ with various Te concentration and $T_C =10 - 18$\,K. They observed approximate scaling between both gaps and $T_C$ with the BCS-ratio for the large gap $4.9 - 7.2$. Taking into account well-established synthesis of qualitative single crystals \cite{Wirth}, the significant data dispersion arose from literature \cite{Park,Kato,Wu,Nakayama,Khasanov,Hanaguri,Miao,Hu,Yin} may originate from, in particular, whether each technique is applicable to Fe(Se,Te), or its superconducting properties sensitivity to crystal surface. In this sense, exemplary could be a combination of a bulk and a local probe with one and the same crystal. In such a comprehensive study, we eliminate the influence of degraded surface, and get a self-consistent data on the order parameter. Here we present pioneer local measurements using intrinsic multiple Andreev reflection effect (IMARE) spectroscopy, and bulk measurements of lower critical field to study superconducting properties of FeSe$_{0.5}$Te$_{0.5}$ single crystals with critical temperature $T_C = 14$\,K. We show that two-band model is applicable to describe superconducting state of FeSe$_{0.5}$Te$_{0.5}$, and directly determine two superconducting gaps and their temperature dependences. In frames of the two-band model, we estimate electron-boson coupling constants, intra- and interband interaction of the two effective bands.

\section{Experimental}
Single crystals of Fe(Te,Se) were prepared using the CsCl/KCl/NaCl eutectic mixture in evacuated quartz ampoules in permanent gradient of temperature. Elongated ampoules with the Fe(Te,Se) charge and CsCl/KCl/NaCl were placed in a furnace so as to maintain their hot end at a temperature of 550$^{\circ}$C and the cold end at a temperature a fortiorti lower than the melting temperature of the salt mixture (480$^{\circ}$C). The chalcogenide charge is gradually dissolved in the hot end of the ampoule and precipitates in the form of single crystals near the cold end. After keeping for 3 weeks in the furnace, iron monochalcogenide plate-like crystals were found directly at the salt melt–solid interface \cite{Chareev1,Chareev2}. The chemical composition of the crystals was determined using a Tescan Vega II XMU scanning electron microscope equipped with an INCA Energy 450 energy-dispersive spectrometer; the accelerating voltage was 20\,kV. Figure 1a shows X-ray diffraction spectrum demonstrating (00l) peaks barely. The crystals grow layer-by-layer (001) with a minor variation of Se:Te content. The lattice parameters are $a = 3.7890$\,{\AA}, $c = 6.0219$\,{\AA}. Noteworthily, the $a$ parameter less varies in (001) plane as compared to the $c$ parameter, resulting in a biaxial stress and additional tetragonal lattice deformation. The latter are seemed to be a reason of enhanced $T_C$ of Te-substituted FeSe as compared to pure composition. The crystals grown were up to 3\,mm in size (electron microscope image of a resulting FeSe$_{0.5}$Te$_{0.5}$ single crystal is shown in Fig. 1c), and possess a single superconducting phase with critical temperature $T_C \approx 14$\,K (Fig. 1a,b,c).

\begin{figure}[h]
	\includegraphics[width=0.5\textwidth]{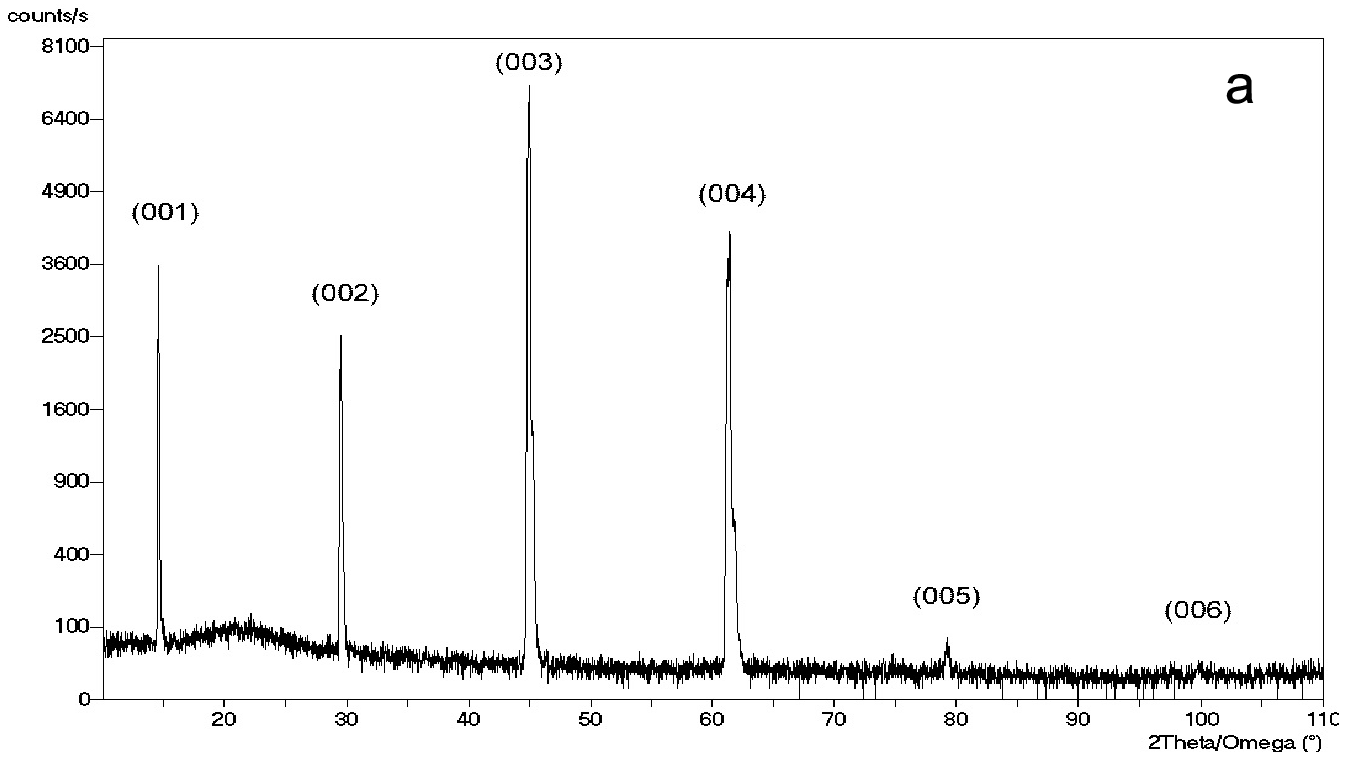}
\includegraphics[width=0.5\textwidth]{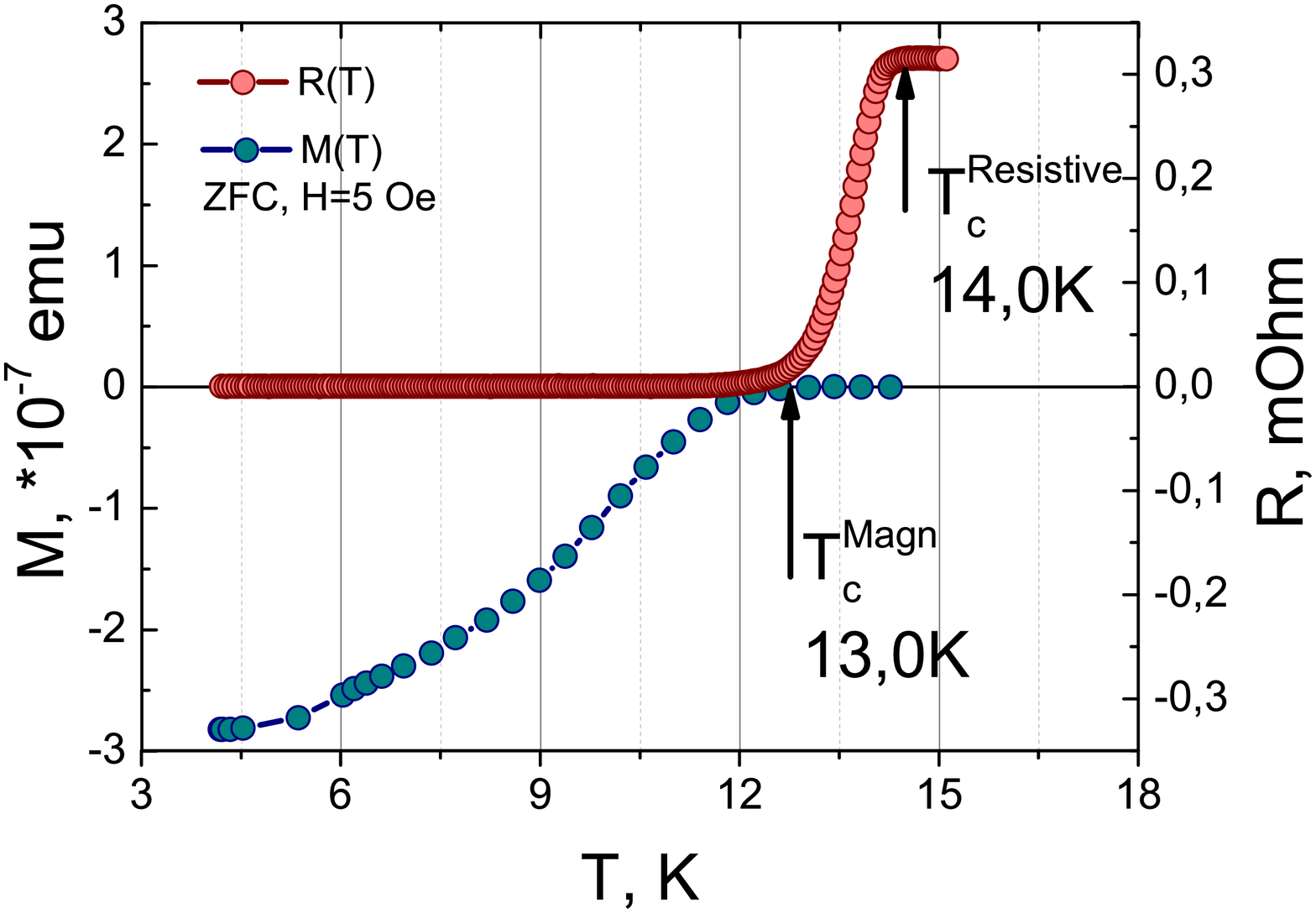}
\includegraphics[width=0.5\textwidth]{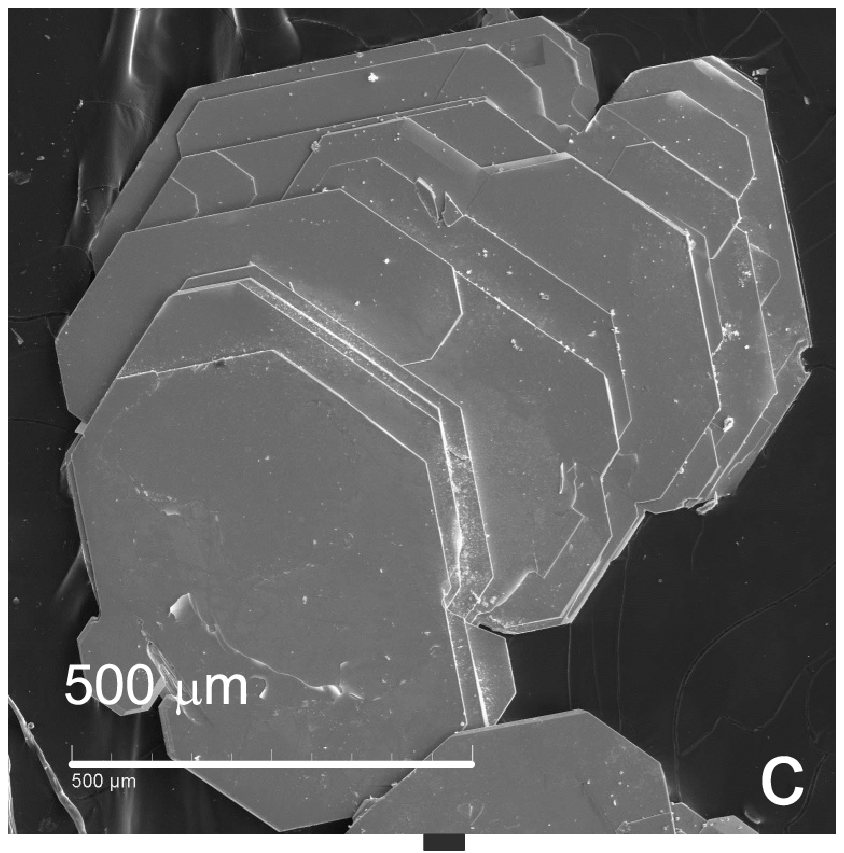}
	\caption{Fig. 1. a) X-ray diffraction spectrum of FeSe$_{0.5}$Te$_{0.5}$ single crystal. b) Resistive (upper panel) and magnetic superconducting transition (lower panel). Arrows indicate the onset of the transitions. c) Electron microscope image of the crystal.}
\end{figure}

Multiple Andreev reflections effect (MARE) spectroscopy is a powerful local probe of bulk superconducting properties. The effect occurs in ballistic SnS (superconductor - normal metal - superconductor) contact which is narrower than carrier mean free path \cite{Sharvin}, and causes a pronounced excess current (``foot'') near zero-bias region in current-voltage characteristic (CVC), and a sequence of dynamic conductance dips (in case of transparency of NS interfaces as high as 95 - 98\,\%)
at certain positions $V_n = 2\Delta/en$, $n$ is natural subharmonic order, called subharmonic gap structure (SGS) \cite{OTBK,Arnold,Averin,Kummel}. Obviously, in two-gap superconductor, two sets of $dI(V)/dV$ features would be observed. The formula for SGS positions is valid at any temperatures up to $T_C$ \cite{OTBK,Kummel}, providing an unique possibility of direct measurements of $\Delta(T)$.

In Andreev spectroscopy studies, we formed symmetrical SnS junctions using a ``break-junction'' technique \cite{Moreland}. Our set-up configuration is detailed in \cite{BJ}. A plate-like single crystal was mounted onto a springy sample holder parallel to crystallographic $ab$-plane, using four In-Ga pads (4-probe circuit) and cooled down to $T = 4.2$\,K. Then, under a gentle mechanical curving of the sample holder the crystal was cracked with a formation of two cryogenic surfaces and a weak link. Since its layered crystal structure, Fe(Se,Te) demonstrates steps and terraces on cryogenic clefts. In our studies, the curving is slight enough to exfoliate the crystal along the $ab$-planes with sliding cryogenic clefts, rather than separate them to a moderate distance. The crack is kept in the bulk of the sample, thus making clean cryogenic clefts and preventing their degradation \cite{BJ}. Sometimes, steps and terraces deliver an SnSn-\dots -S array typical for the ``break-junction'' technique barely. Herewith, studying of such arrays is favourable for many reasons, for example, one surely probes bulk properties of the material \cite{BJ}. The array representing a sequence of $m$ identical SnS junctions, demonstrates intrinsic MARE (IMARE) causing SGS at positions scaled with natural number $m$: $V_n = 2\Delta \times m/en$. Similar to intrinsic Josephson effect \cite{Pon_IJE}, IMARE was observed in all layered superconductors (for a review, see \cite{BJ} and refs. therein). The actual number $m$ could be determined when normalize the $dI(V)/dV$ of array to that of single junction \cite{EPL,BJ}. Usage of IMARE spectroscopy is preferable since being a direct local probe providing a high accuracy of $2\Delta/k_BT_C$ value \cite{BJ}. With it, the break-junction provides a local study of the order parameter within the contact area of about $10 - 30$\,nm \cite{BJ}.

Another important opportunity of the break-junction is a mechanical readjustment of the junction configuration. Precisely tuning the mechanical pressure to the sample holder, it is possible to both form several tens of single and array contacts, and reversibly vary contact diameter and resistance, to collect data statistics and check its reproducibility \cite{BJ}.

Lower critical field $H_{c1}(T)$ and $M(T)$ experiments were made on SQUID magnetometer with sensitivity as high as 10 - 9\,emu (in fields up to 100\,Oe). The magnetic superconducting transition is shown of Fig. 1b (lower panel). Magnetic transition starts right at the point, where resistive transition (upper panel) reaches zero resistivity. For all magnetic measurements the sample was oriented with $H$ parallel to $c$-axis.

\section{Experimental Results: Determination of Superconducting Gaps}

\subsection{Intrinsic Multiple Andreev Reflection Effect (IMARE) Spectroscopy}
Figure 2 shows dynamic conductance spectra (a) and current-voltage characteristics (b, I(V) are shown by corresponding colour) of several Andreev arrays formed at $T = 4.2$\,K. The pronounced foot area with a significant excess current at low biases proves a high-transparency Andreev regime. The $dI(V)/dV$ demonstrate two independent series of conductance dips which obey the SGS formula. For the large gap, the subharmonics are located at $V_1 \approx 5.5 - 6.8$\,mV, $V_2 \approx 3.4$\,mV, and $V_3 \approx 2.3$\,mV, and are interpreted as the first, second and third SGS dips. The first Andreev feature could be shifted towards zero for several reasons \cite{Averin,Kummel,BJ}; if it is the case (see Fig. 2a,b), the gap value is determined using the positions of high-order subharmonics. According to SGS formula, we directly get the large gap $\Delta_L \approx 3.4$\,meV. The positions of features located at $\approx 1.9$\,meV and at $\approx 1$\,meV (resolved in the two spectra in Fig. 2a) do not satisfy the expected positions of 4th and 5th subharmonics of the large gap. These features are possibly the first and the second dips of the small gap $\Delta_S \approx 1$\,meV. The $I(V)$ and $dI(V)/dV$ were scaled with the corresponding numbers $m$ of junctions in the arrays, i.e. normalized to a single junction. The observed two series of gap features agree with the SGS formula and provide two linear dependences between their positions $V_n$ and the inverse number $n$ shown in Fig. 2c. Regardless to the number $m$, the contact dimension and resistance, the positions of the two SGS are reproducible.

\begin{figure}
\includegraphics[width=.5\textwidth]{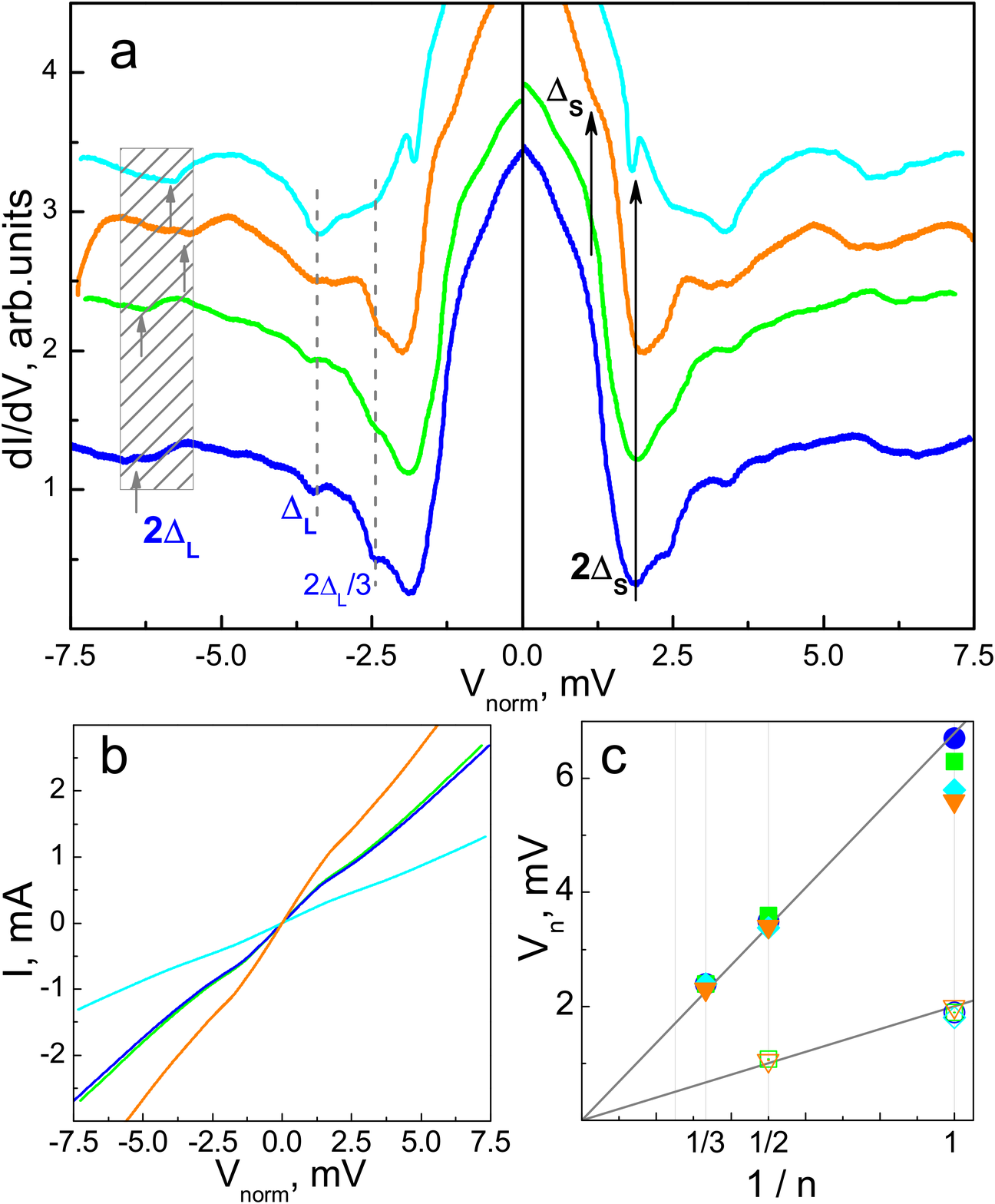}
\caption{Fig.2. a) Normalized dynamic conductance spectra of SnS-Andreev arrays formed in FeSe$_{0.5}$Te$_{0.5}$ single crystal with $T_C \approx 14$\,K. The main features for the large gap corresponding to $2\Delta_L/e$ are marked by dashed rectangle, the second and the third subharmonics located at $\Delta_L/e$ and $2\Delta_L/3e$, are pointed to by dashed vertical lines. $\Delta_L \approx 3.4$\,meV. The Andreev features possibly related to the small gap $\Delta_S \approx 1$\,meV are marked with black arrows. b) Normalized current-voltage characteristics for the above contacts. c) The dependence of the positions of Andreev features versus the inverse subharmonic order, $1/n$ for the above contacts. Gray lines are guidelines.}
\end{figure}

\subsection{Lower Critical Field Measurements}
While the break-junction experiments provide information on microcontacts, it is crucial to confirm the results with a bulk probe. London penetration depth $\lambda$ is a fundamental parameter, which characterizes the SC condensate and probes the gap structure of bulk crystal. Its temperature dependence is determined by the gap function $\Delta(T)$: $\lambda(T) = \lambda(0) + \delta\lambda(T)$, where $\delta\lambda(T) \propto \exp(-\delta k_BT)$ at low $T$ for a nodeless superconducting gap with $s$-wave symmetry. We estimated the penetration depth using the traditional Ginzburg-Landau (GL) theory, where lower critical field $H_{c1}$ (for $H \parallel c$ geometry) is given by $\mu_0H_{c1} = (\phi_0/4\pi \lambda_{ab}^{-2}){\rm ln}\kappa_c$, $\phi_0$ is the magnetic flux quantum and $\kappa_c$ is the Ginzburg-Landau parameter.


We used a well-known approach to determine lower critical field by measuring the magnetization M as a function of H and then identified the deviation of the linear Meissner response which would correspond to the vortex penetration. Here we assume that no surface barriers are present, thus assuring that $H_{c1}$ coincides with vortex penetration \cite{Mahm2,Mahm1}.

Generally speaking, penetration of first vortex may not reflect the actual reaching of lower critical field, if surface barrier is present \cite{Bean}. However, as it was shown for several types of 11 family single crystal materials (FeSe --- \cite{Mahm2}, FeS, FeSeS --- \cite{HystFESES}, FeSeTe --- \cite{HystFESETE1,HystFESETE2} magnetic hysteresis loops are symmetric close to the critical temperatures of these materials, which means the absence of surface barriers and thus validating the determination of $H_{c1}(T)$ by means of deviation from linear Meissner behaviour criterium.

Since magnetization $M(H)$ shows the linear behavior (see Fig. 3a) in the region below $H_{c1}$ and trapped magnetization above $H_{c1}$ may be represented as second power polynomial the magnetization $M(H)$ may be written as:

\begin{eqnarray}
\label{ModelHc1}
M(H)&=& aH+b \nonumber \qquad\qquad\qquad\quad \text{\rm for } H<H^*\\
M(H)&=& aH+b +c (H-H^*)^2 \quad \text{\rm for } H>H^*.
\end{eqnarray}

For every data point $H_i$ we take
$H^*=H_i$
and find the best fitting parameters and calculate the correlation index (coefficient of determination) for this curve (Fig. 3b). The maximum of correlation index gives us lower critical field $H_{c1}$.

\begin{figure}[h]
	\includegraphics[width=0.5\textwidth]{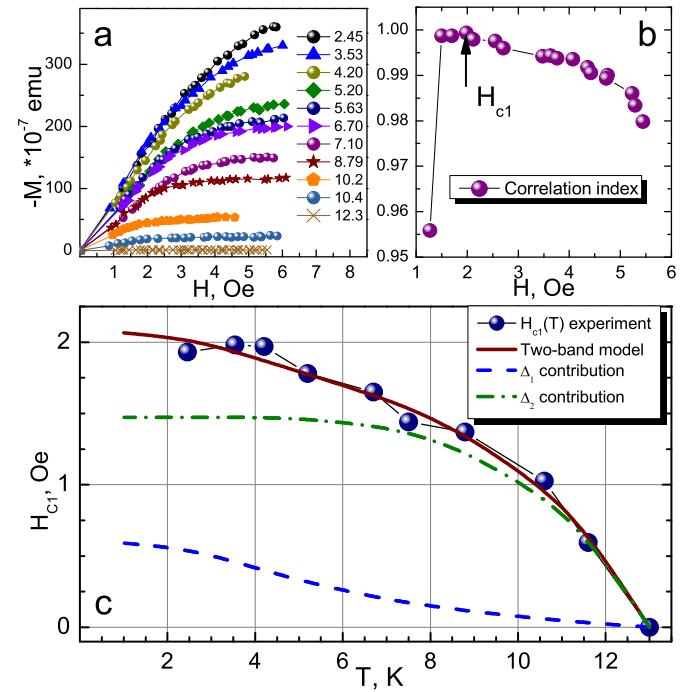}
	\caption{Fig. 3. a) Temperature dependence of the first critical field $H_{c1}(T)$ (circles) fitted using two-band model (solid line). The partial contributions of the band with the large and the small gap are shown by dashed lines. b) A set of magnetization curves for several temperatures. $H_{c1}$ Was determined at the point, where the correlation index reaches its maximum value. c) Correlation index for on of temperatures.}
	\label{hc1all}
\end{figure}

Figure 3c shows the temperature dependence $H_{c1}(T)$. For describing the experimental data we apply the so called two-band $\alpha$-model \cite{carrington_2003}:
\begin{equation}
\tilde{\rho_s} (T)= \varphi_1 \tilde{\rho}_{s1}(T)+\varphi_2 \tilde{\rho}_{s2} (T)
\end{equation}
This model considers a normalized superfluid density for the superconductor having two independent condensates with a normalized superfluid densities $\rho_{s1}$ and $\rho_{s2}$ in the first and second band respectively, taken with weighting factors $\varphi_1$ and $\varphi_2=1-\varphi_1$. More detailed take on this two-band $\alpha$-model can be found in our previous works \cite{MahmFeSeS,MahmKuzm} or in other works \cite{Mahm2}.

Here we used this model to confirm the IMARE data. We take the gap values as 1 and 3.3 meV that corresponds to $\alpha_{1}=1.6$ and  $\alpha_{2}=5.5$ and fitted $H_{c1}$ using two free parameters. As shown in Fig. 3c, this approach gives very good agreement between experimental and model-fitting data.

\section{Discussion}
The magnitude of the two superconducting gaps, $\Delta_L \approx 3.3 - 3.4$\,meV and $\Delta_S \approx 1$\,meV, determined using a bulk probe (lower critical field) and a local probe (IMARE spectroscopy) is reproducible. This demonstrates both the bulk nature of the two-gap superconductivity in FeSe$_{0.5}$Te$_{0.5}$, and high homogeneity of the crystal grown.

\begin{figure}
\includegraphics[width=.5\textwidth]{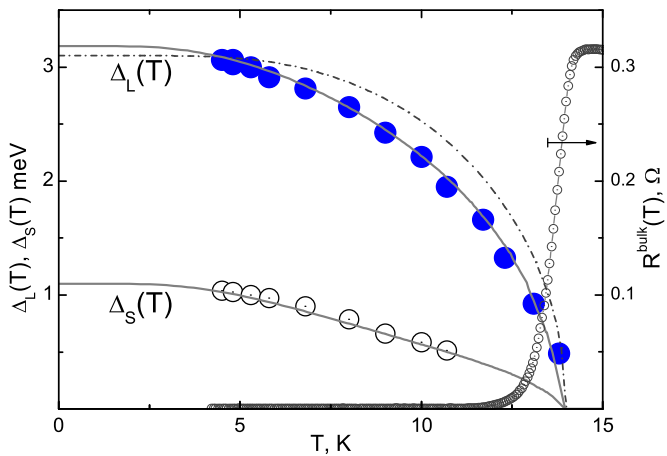}
\caption{Fig. 4. Temperature dependence of the large gap (blue circles) and the small gap (open black circles). The experimental data are fitted using two-band BCS-like model with a renormalized BCS-integral, the fits are shown by gray lines. Single-band BCS-like curve (dash-dot line), and bulk resistive transition of the sample (connected circles) are presented for comparison.}
\end{figure}

Temperature dependence of two superconducting gaps directly obtained in IMARE experiment is shown in Fig. 4. The $\Delta_L(T)$ is slightly curved as compared to the single-band behaviour (shown by dash-dot line in Fig. 4), whereas the $\Delta_S(T)$ deviates from the BCS-type much stronger. Obviously, a single-band model is insufficient to describe the experimental $\Delta_{L,S}(T)$. The two gaps turn to zero at common critical temperature $T_C^{local} \approx 14$\,K which corresponds to the contact area (of $10 - 30$\,nm in size) transition to a normal state. Generally, $T_C^{local}$ is not directly related to $T_C$ determined in bulk probes (see Fig. 1b). In the case, $T_C^{local}$ nearly matches the onset of the R(T) transition (shown in Fig. 4 by connected gray circles). Being not averaged over the bulk of the sample, $T_C^{local}$ facilitates estimation of true local BCS-ratio. For the large gap, we obtain $2\Delta_L/k_BT_C^{local} \approx 5.6$ strongly exceeding the BCS-limit 3.5 thus supporting a strong coupling in ``driving'' bands where the large gap is developed. For the small gap, $2\Delta_S/k_BT_C^{local} \approx 1.7 < 3.5$ and originates from a nonzero interband interaction.

\begin{figure}
\includegraphics[width=.45\textwidth]{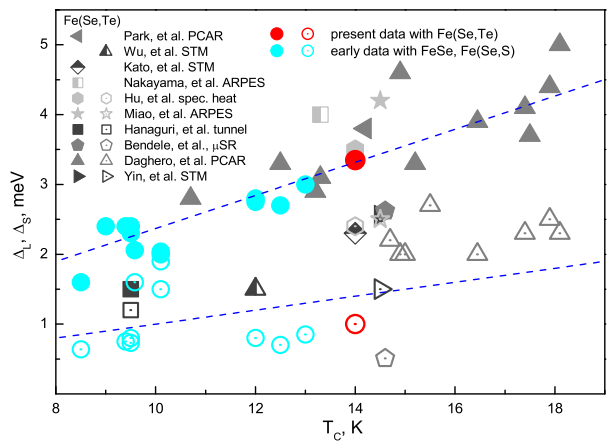}
\caption{Fig. 5. The large (solid symbols) and the small gap (open symbols) dependence on critical temperature $T_C$ for iron selenide superconductors: the data of current work (red circles), the data with Te-substituted crystals from literature \cite{Park,Kato,Wu,Nakayama,Daghero,Khasanov,Hanaguri,Hu,Miao,Yin} (gray symbols), and our earlier data with pure FeSe and Fe(Se,S) from \cite{FeSe,FeSe2013,UFN,MahmFeSeS,Chareev2013} (cyan circles). Dashed lines are guidelines.}
\end{figure}

Turning to literature \cite{Park,Kato,Wu,Nakayama,Daghero,Khasanov,Hanaguri,Hu,Miao,Yin}, in Fig. 5 we summarize the values of the large and the small superconducting gaps in dependence of critical temperature. $\Delta_L$ and $\Delta_S$ roughly scale with $T_C$, with the determined here gap values dispose right the middle of the data point range. The $2\Delta_L/k_BT_C$ is similar to those obtained in PCAR \cite{Daghero} and specific heat \cite{Hu} studies with Fe(Se,Te), and to those reported by us earlier, in IMARE measurements of both single crystal \cite{Chareev2013,UFN} and polycrystalline pure FeSe \cite{FeSe,FeSe2013,UFN}, and in specific heat measurements of S-substituted iron selenide \cite{MahmFeSeS}. The linear $\Delta(T_C)$ tendency favors a minor influence of (Se,Te,S) substitution on the fundamental pairing.

Despite the majority of theoretical studies consider three or more interacting bands \cite{Maiti,Kontani,Mazin}, a two-band model adequately fits our experimental data obtained by the two techniques. Therefore, we can consider two effective bands to estimate electron-boson coupling constants $\lambda_{ij} = V_{ij}N_j$ ($V$ is interaction matrix element, $N$ is normal carrier density of states (DOS) at the Fermi level, $i,j = 1,2$, hereafter the effective band with the large gap is labelled as ``1'', that with the small gap --- as ``2'') and to make some important conclusions concerning the two-gap superconductivity in FeSe$_{0.5}$Te$_{0.5}$. We used simple two-band model based on Moskalenko and Suhl system of equations \cite{Mosk,Suhl} with a renormalized BCS integral to describe the experimental data in Fig. 5. As fitting parameters, we take: the Debye energy $\omega_D = 25$\,meV \cite{Wen}, the DOS ratio $\alpha \equiv \lambda_{12}/\lambda_{21} = N_2/N_1$ and the ratio between intra- and interband effective couplings $\beta = \sqrt{V_1V_2}/V_{12}$. Since BCS-based equations by Moskalenko and Suhl manage with pure pairing constants $\lambda_{ij} = \lambda_{ij}^{Full} - \mu_{ij}^{\ast}$ ($\mu_{ij}^{\ast}$ is a set of Coulomb repulsion constants), which corresponds to the case of weak BCS-coupling. In order to restore strong-coupling constants $\lambda_{ij}^{Full}$, one should add $\mu_{ij}^{\ast}$, if any, to $\lambda_{ij}^{Fit}$ estimated during the fitting procedure. The latter is detailed in \cite{fit,fit2016}. The fits shown by solid lines in Fig. 5 agree with the experimental points. Given zero Coulomb pseudopotentials $\mu^{\ast} = 0$ suggested in \cite{Maiti,Mazin}, we get $\lambda_{11} = 0.3$,  $\lambda_{22} = 0.22$, $\lambda_{12} = 0.125$, $\lambda_{21} = 0.02$, which lead to unrealistic $\alpha = 6$ and $\beta = 5$, seeming to be too high for an iron-based superconductor. When assuming a moderate Coulomb repulsion $\mu^{\ast}_{eff} = 0.2$, one immediately gets $\lambda_{11} = 0.5$,  $\lambda_{22} = 0.42$, $\lambda_{12} = 0.32$, $\lambda_{21} = 0.21$. The latter estimation leads to rather reasonable DOS ratio 1.5, and an effective intraband coupling, 1.7 times stronger than interband one.

\section{Conclusions}
We have studied the two-gap superconducting state of FeSe$_{0.5}$Te$_{0.5}$ single crystals with $T_C \approx 14$\,K. Using a pioneer combination of a bulk probe, lower critical field measurements, and a local probe, IMARE spectroscopy (break-junction technique), we directly determined the two superconducting gaps, $\Delta_L = 3.3 \pm 0.3$\,meV and $\Delta_S = 1.0 \pm 0.1$\,meV, and their temperature dependences. The $H_{c1}$ and IMARE experimental data are well-described in frames of a two-band model. Considering two effective bands, we estimated electron-boson coupling constants, and some other important parameters of superconducting state. The BCS-ratio for the large gap $2\Delta_L/k_BT_C \approx 5.6 > 3.5$ indicates a strong intraband coupling (described by diagonal constants $\lambda_{ii}$, $i = 1,2$), whereas nondiagonal $\lambda_{ij}$ are moderate, about 1.7 times weaker than interband one. For some reasonable value of Coulomb repulsion ($\mu^{\ast} = 0.2$), electron-boson coupling constants were estimated as $\lambda_{11} = 0.5$,  $\lambda_{22} = 0.42$, $\lambda_{12} = 0.32$, $\lambda_{21} = 0.21$.

\section{Acknowledgments}
T.E.K., A.V.S., and A.V.M. were supported by RSF grant 16-12-10507. D.A.C. and E.S.M. thank RFBR support (grant 16-32-00435-mol-a), and Act 211 of the Government of Russian Federation, agreement No. 02.A03.21.0006, and Program of Competitive Growth of Kazan Federal University. The measurements were partly done using research equipment of the Shared Facilities Center at LPI.

\end{document}